\documentstyle[twocolumn,aps]{revtex}

\input psfig.tex

\newcommand{\ltrsim}{\mathrel{\lower .3ex \rlap{$\sim$}
\raise .5ex\hbox{$<$}}}
\newcommand{\gttrsim}{\mathrel{\lower .3ex \rlap{$\sim$}
\raise .5ex\hbox{$>$}}}

\begin{document}
\draft

\twocolumn[\hsize\textwidth\columnwidth\hsize\csname %
@twocolumnfalse\endcsname

\title{
Spin and Charge Excitations in the
Two-Dimensional $t$-$J$ Model: Comparison with Fermi and Bose Systems
}

\author{
W. O. Putikka$^{a,b,c}$, M. U. Luchini$^d$ and R. R. P. Singh$^{a,e}$
}

\address{
$^a$Institute for Theoretical Physics, University of California, Santa Barbara, CA 93106\\
$^b$Theoretische Physik, ETH H\"onggerberg, CH-8093 Z\"urich, Switzerland\\
$^c$Department of Physics, The Ohio State University, Mansfield, OH 44906$^*$\\
$^d$Department of Mathematics, Imperial College, London SW7 2BZ, United
Kingdom\\
$^e$Department of Physics, University of California, Davis, CA 95616$^*$\bigskip\\
}

\maketitle
\begin{abstract}
Using high temperature series we calculate temperature derivatives of the 
spin-spin and density-density correlation functions to investigate the low
energy spin and charge excitations of the two-dimensional 
$t$-$J$ model.  We find that the temperature derivatives indicate different
momentum dependences for the low energy spin and charge excitations. 
By comparing short distance density-density correlation functions with those of
spinless fermions and hard core bosons, we find that the $t$-$J$ model
results are intermediate between the two cases, being closer to those of
hard core bosons. The implications of these results for superconductivity
are discussed.
\vspace{0.3in}
\end{abstract}
]

The nature of the ground state and low energy excitations for 
two-dimensional strongly correlated electrons doped slightly  away from
half-filling is of considerable interest for understanding the properties
of high temperature superconductors\cite{rice}.  
The $t$-$J$ model on a square lattice
is a widely studied model used to investigate these problems.  While the
properties of a single hole introduced into an antiferromagnet are well
understood\cite{lee}, how to extend these results to a finite density of holes 
remains a subject of much current research \cite{numerics}.

For conventional metals, the electron spectral function is the simplest
way to investigate the energy and momentum dependence of the single particle
excitations\cite{mahan}.  With high temperature series we cannot easily calculate the
spectral function of the 2D $t$-$J$ model directly.
From the high temperature series for the momentum distribution $n_{\bf k}$
we calculated\cite{putikka1} the temperature derivative $dn_{\bf k}/dT$ which we used as a
proxy for the momentum dependence of the low energy part of the spectral
function.
Our results for $dn_{\bf k}/dT$
showed that the low energy excitations of the 2D $t$-$J$ model are spread
throughout the Brillouin zone and are in general
not conventional quasiparticles.  A consequence of this
result is that $dn_{\bf k}/dT$ does not completely determine
the momentum dependence of the low energy elementary excitations in the 2D $t$-$J$ model.
To further investigate the nature of the low energy elementary excitations, we have
extended our calculations to the equal time spin-spin and density-density 
correlation functions, $S({\bf q})$ and $N({\bf q})$ respectively, and their
temperature derivatives $dS({\bf q})/dT$ and $dN({\bf q})/dT$.

We have calculated high temperature series for $S({\bf q})$ and $N({\bf q})$ of
the 2D $t$-$J$ model to twelfth order in inverse temperature $\beta=1/k_B T$.
Our calculations extend previous series calculations\cite{singh,putikka2} for the correlation
functions of the $t$-$J$ model.  
The $t$-$J$ Hamiltonian is given by
\begin{equation}
H=-tP\sum_{\langle ij\rangle,\sigma}\left(c_{i\sigma}^{\dagger}
c_{j\sigma} + c_{j\sigma}^{\dagger}c_{i\sigma}\right)P+J\sum_{
\langle ij\rangle}{\bf S}_i\cdot{\bf S}_j,
\end{equation}
where the sums are over pairs of nearest neighbor sites and the projection operators
$P$ eliminate from the Hilbert space states with doubly occupied sites.  
The definitions of the spin-spin and density-density correlation functions are
\begin{eqnarray}
S({\bf q}) = \sum_{\bf r}e^{i{\bf q}\cdot{\bf r}}\langle S^z_0 S^z_{\bf r}\rangle\\
N({\bf q}) = \sum_{\bf r}e^{i{\bf q}\cdot{\bf r}}\langle\Delta n_0\Delta n_{\bf r}
\rangle,
\end{eqnarray}
where 
$S^z_{\bf r}={1\over2}\sum_{\alpha\beta}c^{\dagger}_{{\bf r}\alpha}\sigma^z_
{\alpha\beta}c_{{\bf r}\beta}$ and
$\Delta n_{\bf r} = \sum_{\sigma}c^{\dagger}_{{\bf r}\sigma}c_{{\bf r}\sigma}-n$.
The series are extrapolated to $T=0.2J$ by Pad\'e approximants and a ratio
technique used previously\cite{putikka1} for $n_{\bf k}$.

To interpret our results for the $t$-$J$ model we examine $dN({\bf q})/dT$ for
the tight-binding and spinless fermion models.  For these non-interacting models
$dN({\bf q})/dT$ is given by
\begin{equation}
{dN({\bf q})\over dT}=-g\int{d{\bf k}\over(2\pi)^2}\left(n_{\bf k}{dn_{{\bf k}+{\bf q}}
\over dT}+n_{\bf k}{dn_{{\bf k}-{\bf q}}\over dT}\right),
\end{equation}
where $g=2$ for tight binding and $g=1$ for spinless fermions.  
The properties of $dN({\bf q})/dT$ for the non-interacting models are determined
by the convolution of $n_{\bf k}$ with $dn_{\bf k+q}/dT$.  At low temperatures, due
to Fermi statistics $n_{\bf k}$ is large only inside the Fermi surface and
$dn_{\bf k+q}/dT$ is negative just inside the Fermi surface, positive just outside
and close to zero elsewhere.  The convolution will then give a significant contribution
to the integral in Eq. 4 when {\bf q} is such that only one of the positive or
negative parts of $dn_{\bf k+q}/dT$ overlaps $n_{\bf k}$.
This occurs for ${\bf q}\approx 0$ and ${\bf q}\approx 2{\bf k}_F$,
as demonstrated in Fig. 1 where we plot
$dN({\bf q})/dT$ for the tight binding and spinless fermion models.  The main
features of these plots are a large positive spike at ${\bf q}\approx 0$ and a
smaller but more extended negative dip located at ${\bf q}\approx 2{\bf k}_F$.
The shape of the $2{\bf k}_F$ line depends on the nature of the Fermi surface
(hole like or electron like) but in both cases $2{\bf k}_F$ is a continuous
curve in the Brillouin zone.  
\begin{figure}[htb]
\centerline{\psfig{figure=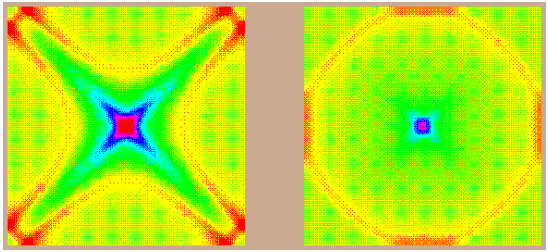,height=1.3in}}
\vspace{0.1in}
\caption{(color)
Full Brillouin zone plots of $dN({\bf q})/dT$ for left, the tight binding model and
right, the spinless fermion model.  The parameters for both models are $n=0.8$ and
$T=0.08t$.  Large, positive central spikes show up as red 
and dark blue, while the negative features at $2{\bf k}_F$ are orange and yellow.
}
\end{figure}

For the $t$-$J$ series calculations the ${\bf q}\approx 0$
(long range) parts of the correlation functions have the least accuracy,
while we expect the correlation functions at larger wavevectors (short range)
to be well determined.  Consequently, we concentrate in our analysis on the
locations and properties of the ``$2{\bf k}_F$'' features in the $t$-$J$ model
correlation functions.  Using the non-interacting models as guides, we search
for the ``$2{\bf k}_F$'' features in the $t$-$J$ correlation functions by looking
for the largest negative values of $dN({\bf q})/dT$ and
$dS({\bf q})/dT$.
\begin{figure}[htb]
\centerline{\psfig{figure=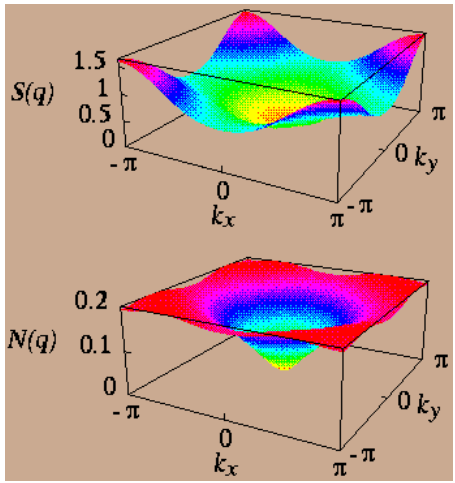,height=2.5in}}
\vspace{0.1in}
\caption{(color)
Three dimensional full Brillouin zone plots of correlation functions for the $t$-$J$ model.
Top, $S({\bf q})$ and bottom, $N({\bf q})$.  The parameters for both plots are $n=0.8$,
$J/t=0.4$ and $T=0.2J$.
}
\end{figure}

Results for the $t$-$J$ model $N({\bf q})$ and $S({\bf q})$ with electron density $n=0.8$,
$J/t=0.4$ and $T=0.2J$ are shown in Fig. 2.  Our data are in good agreement with previous
calculations at higher temperatures\cite{singh,putikka2}.  
The Brillouin zone sums of the correlation functions
agree with their respective sum rules to within 0.5\%.  Using data at $T=0.2J$ and $T=0.4J$
we calculate $\Delta N({\bf q})/\Delta T$ and $\Delta S({\bf q})/\Delta T$ as approximations
for the temperature derivatives at $\bar{T}=0.3J$.
\begin{figure}[htb]
\centerline{\psfig{figure=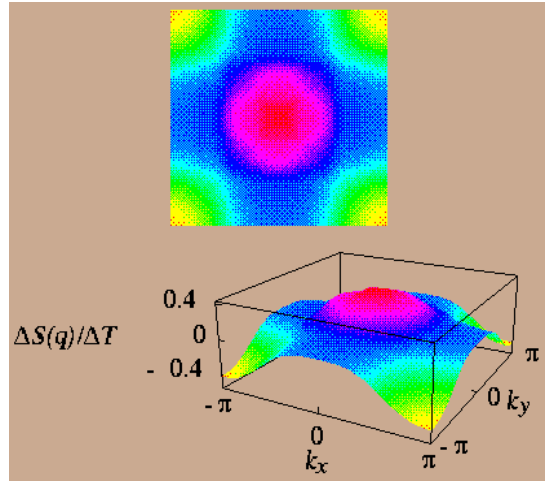,height=2.5in}}
\vspace{0.1in}
\caption{(color)
Two views of the full Brillouin zone plot of $\Delta S({\bf q})/\Delta T$ in units of $J^{-1}$
for the $t$-$J$ model.  The data
and color coding are the same for both plots, with parameters $n=0.8$, $J/t=0.4$ and
$\bar{T}=0.3J$.
}
\end{figure}
\begin{figure}[htb]
\centerline{\psfig{figure=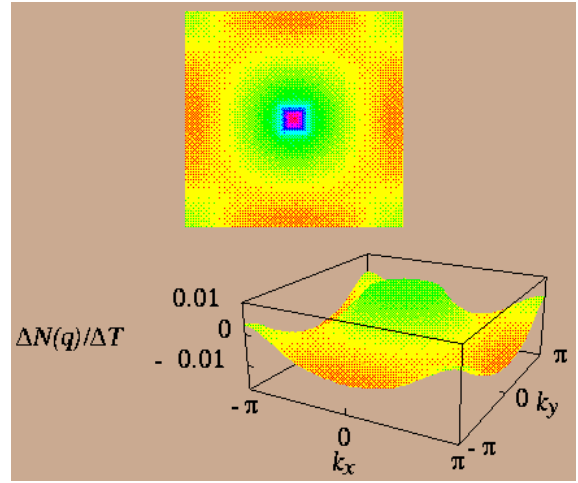,height=2.5in}}
\vspace{0.1in}
\caption{(color)
Two views of the full Brillouin zone plot of $\Delta N({\bf q})/\Delta T$ in units of $J^{-1}$
for the $t$-$J$ model.  The central peak with a maximum value of $0.08J^{-1}$
has been truncated at $0.01J^{-1}$ in the lower plot to better show the behavior at large 
wavevectors.  The data and color coding are the same for both plots, with parameters $n=0.8$, 
$J/t=0.4$ and $\bar{T}=0.3J$.
}
\end{figure}

Results for $\Delta S({\bf q})/\Delta T$ are shown in Fig. 3.  The peak at ${\bf q}\approx 0$
is considerably broader than for the non-interacting models, in agreement with the temperature
dependence observed in previous calculations\cite{singh}.  The only part of the Brillouin zone where
$\Delta S({\bf q})/\Delta T$ is negative is a roughly circular region of approximate radius $0.6\pi$
centered on ($\pi$, $\pi$).  
We note that even at the lowest temperature accesible to us, the spin correlations
are still peaked at ($\pi,\pi$). The correlation length, around ($\pi,\pi$) appears to be
saturating or perhaps even decreasing with lowering $T$ \cite{singh}. This could be
taken as an indication for incommensurate correlations at still lower temperatures, as
the peak at ($\pi$, $\pi$) splits into several distinct peaks.
Experimentally, for the high temperature superconducting materials,
incommensurate spin-correlations only arise below about $100$ K \cite{aeppli}.

The negative feature in $\Delta S({\bf q})/\Delta T$ does not form a closed curve in the
Brillouin zone.  In particular, $\Delta S({\bf q})/\Delta T$ remains positive from (0, 0) to
($\pi$, 0), with no indication of a Fermi surface in the low energy spin excitations along 
this line.  Interpreting part of the negative feature in $\Delta S({\bf q})/\Delta T$ as due
to an underlying momentum distribution of itinerant spin degrees of freedom gives disconnected
arcs of low energy spin excitations centered near ($\pi/2$, $\pi/2$) and extending perpendicular
to the zone diagonals.  
These features are located near the peaks observed\cite{putikka1} in $dn_{\bf k}/dT$ and 
$|{\bf \nabla}_{\bf k}n_{\bf k}|$,
consistent with the strongest features in $n_{\bf k}$ being due to an 
underlying  spinon Fermi surface.

\begin{figure}[htb]
\psfig{figure=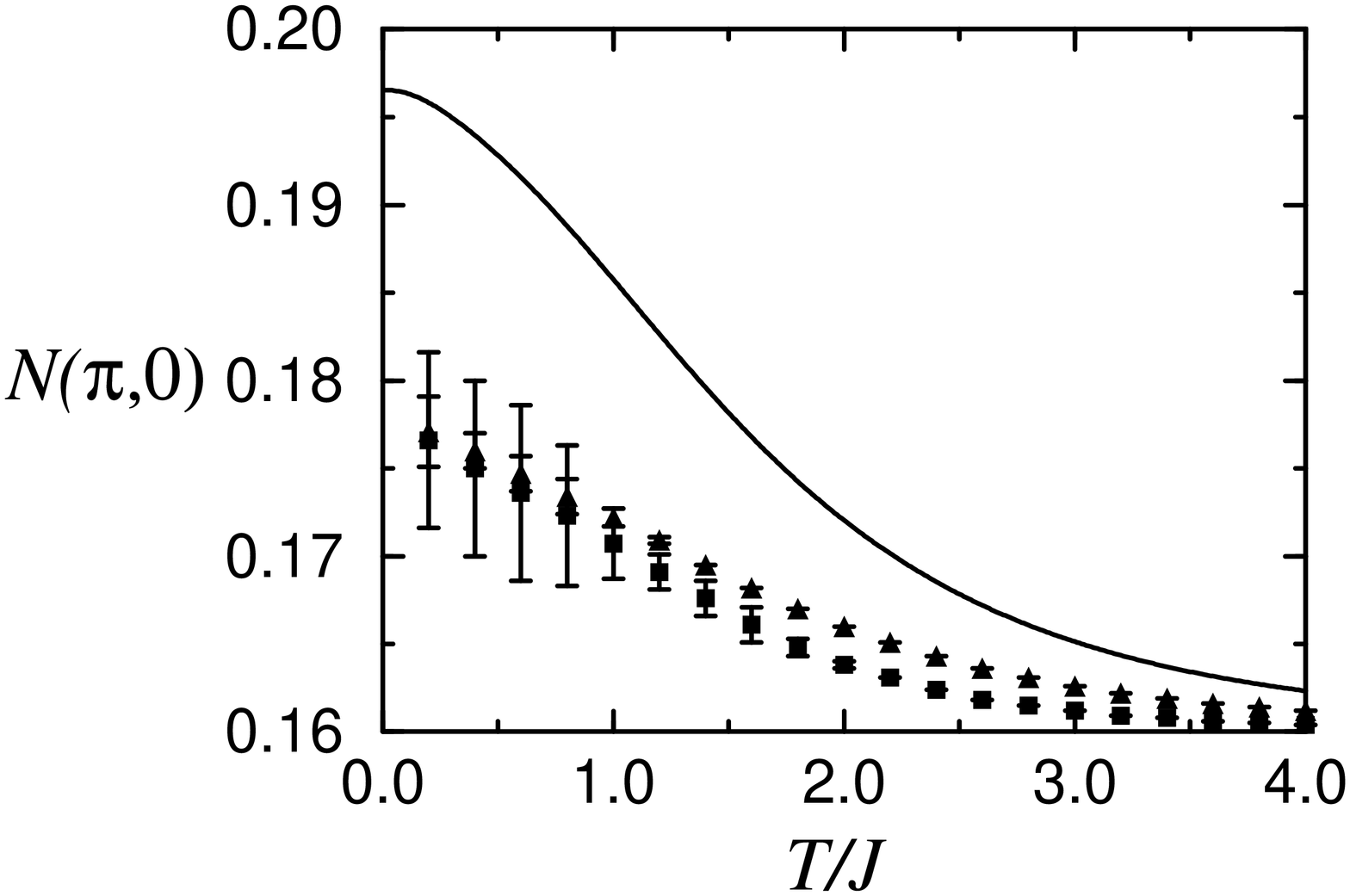,height=2in,angle=-90}\vspace{0.2in}
\psfig{figure=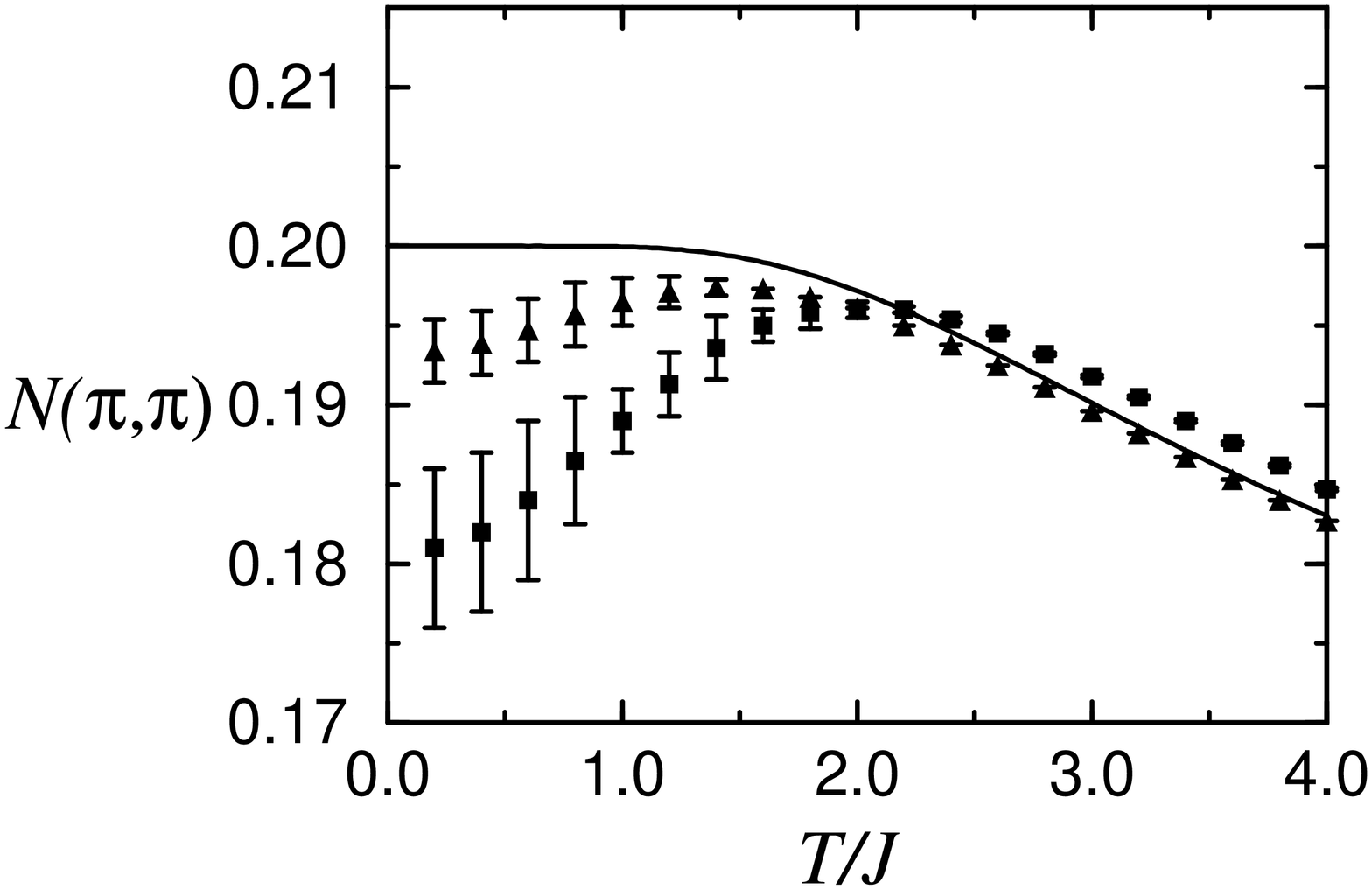,height=2in,angle=-90}
\vspace{0.1in}
\caption{
Top, $N$($\pi$, 0) and bottom, $N$($\pi$, $\pi$) versus temperature.  
Solid line: spinless fermions, squares: hard core bosons
and triangles: $t$-$J$ model.  The density is $n=0.8$ for all of the data sets and $J/t=0.4$
for the $t$-$J$ model.
}
\end{figure}

Results for $\Delta N({\bf q})/\Delta T$ are shown in Fig. 4.  The peak for ${\bf q}\approx 0$ is
much sharper than for $\Delta S({\bf q})/\Delta T$ and more like the non-interacting models.  The
negative feature in $\Delta N({\bf q})/\Delta T$ does make a closed curve in the Brillouin zone,
in contrast to $\Delta S({\bf q})/\Delta T$.  The location and shape of the negative feature in
$\Delta N({\bf q})/\Delta T$ are similar to the $2{\bf k}_F$ line in $dN({\bf q})/dT$ of the
spinless fermion model at the same density.  The similarity extends to having the strongest negative
feature in both models near ($\pi$, 0).  
In the spinless fermion model this is due to parts of the $2{\bf k}_F$ curve overlapping 
after being translated back into the first zone.
The momentum width of the negative feature in $\Delta N({\bf q})/\Delta T$ for the
$t$-$J$ model is considerably broader than the $2{\bf k}_F$ line for spinless fermions and this width
is not temperature dependent down to $T=0.2J$.  Interpreting the negative feature in 
$\Delta N({\bf q})/\Delta T$ for the $t$-$J$ model as arising from an underlying momentum distribution
for the charge degrees of freedom gives low energy charge excitations smeared out over a range of
momenta near ${\bf k}_F$ of spinless fermions at the same density.  For $n=0.8$ the charge
excitations are centered on the zone diagonals and away from ($\pi$, 0).

The low temperature momentum dependence of $N({\bf q})$ for the $t$-$J$ model, while in general similar
to spinless fermions\cite{putikka2}, has important differences from spinless fermions 
at large momenta.  As shown
in Fig. 2, $N({\bf q})$ for the $t$-$J$ model near ($\pi$, $\pi$) and ($\pi$, 0) is slightly
smaller than $0.2$, the value of $N({\bf q})$ for spinless fermions when ${\bf q}>2{\bf k}_F$.
This means that holes in the $t$-$J$ model have
a greater tendency to be nearest or next-nearest neighbors than in the spinless fermion model, with
the effect largest for nearest neighbors.  This tendency has also been observed in 
exact diagonalization\cite{poilblanc}
and Green's function Monte Carlo calculations\cite{sorella}, in both of which 
the effect is much larger than observed in the series calculation.  This enhancement is
probably due to finite size effects, though more work is needed to fully understand the differences
between the series calculations and the Green's function Monte Carlo and exact diagonalization
calculations.
\begin{figure}[htb]
\centerline{\psfig{figure=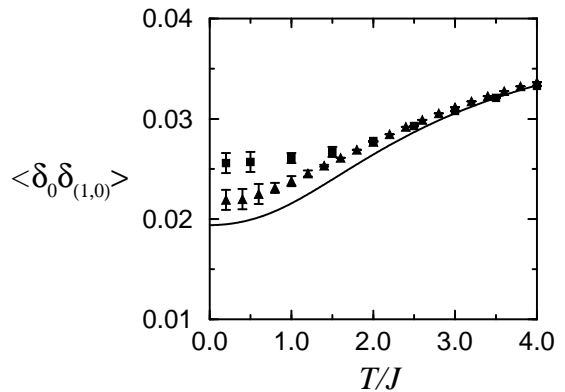,height=2in,angle=-90}}
\vspace{0.1in}
\caption{
Real space, nearest neighbor hole-hole correlation functions.  Solid line: spinless fermions,
squares: hard core bosons and triangles: $t$-$J$ model.  The density is $n=0.8$ for all three
data sets and $J/t=0.4$ for the $t$-$J$ model.
}
\end{figure}

Fig. 5 shows the temperature dependence of $N({\bf q})$ at ($\pi$, 0) and
($\pi$, $\pi$) for the $t$-$J$, spinless
fermion and hard core boson models.  All of these models have an infinite on-site repulsion
which sets the overall scale for $N({\bf q})$.  
The hard core boson results shown in Figs. 5 and 6 are derived from a twelfth order high temperature
series for $N({\bf q})$ of hard core bosons.  The
ground states of spinless fermions and hard core bosons are known: Fermi sea
with a well defined Fermi surface for spinless fermions
and a superfluid for hard core bosons.  The data for the $t$-$J$ model lie between these
two cases, leaving open the possibility that the $t$-$J$ model has a superfluid ground state.
The real space nearest neighbor hole-hole correlation function $\langle\delta_0\delta_{(1,0)}\rangle$
further supports this behavior.  Fig. 6 shows the temperature dependence of 
$\langle\delta_0\delta_{(1,0)}\rangle$ for the $t$-$J$, spinless fermion and hard core boson
models.  Again, at low temperatures the $t$-$J$ data is between the spinless fermion and hard core
boson results.

The temperature dependences shown in Figs. 5 and 6 are also interesting.  At high temperatures all three
models have similar values, with the $t$-$J$ data very close to spinless fermions.  As the
temperature is lowered below $T\ltrsim 1.5J$ (for $J/t=0.4$) the $t$-$J$ data deviates from spinless
fermions towards hard core bosons.  This temperature scale is too high to be due to coherent spin
fluctuations.  Also, the $t$-$J$ results are only weakly dependent on $J/t$ for $J/t\ltrsim 0.5$ and
persist to $J/t=0$.  This shows that the enhancement of $\langle\delta_0\delta_{(1,0)}\rangle$
in the $t$-$J$ model relative to spinless fermions is due to the presence of two spin species, but not
due to a direct spin interaction.

For larger $J/t$, in the phase separation parameter regime, the nearest neighbor density
correlation grows rapidly as the temperature is lowered. This latter increase is clearly
due to an attraction between the holes mediated by $J$. In contrast to this, the high
temperature devitaion from spinless fermions towards hard core bosons appears to be
a statistical effect. This is, perhaps, analogous to the Hanbury-Brown and Twiss correlation
well known for quantum particles \cite{Baskaran,Baym}.
Thus it appears that strong on-site repulsion tends to
make the charge degrees of freedom in the $t$-$J$ model similar to
hard core bosons at a high temperature scale\cite{holons}.  
This would suggest a superfluid ground state for the model, where
the spin degrees of freedom 
merely help to choose the symmetry of the superfluid state.
For antiferromagnetic spin correlations d-wave might be favored, while for ferromagnetic
spin correlations p-wave symmetry could be favored.  These ideas find some support in two-hole
calculations\cite{hamer}, 
but require considerable further investigation.

In conclusion, from series calculations we find that $\Delta S({\bf q})/\Delta T$ and $\Delta N({\bf q})/\Delta T$ for the 2D
$t$-$J$ model are quite different, giving further support to non-quasiparticle elementary excitations
in 2D strongly correlated systems.  Interpreting these results as due to itinerant degrees of freedom
with underlying momentum distributions gives low energy spin and charge excitations near the zone
diagonals, but with different momentum dependences, and absent near ($\pi$, 0).  The
charge correlations show considerable similarity with hard-core bosons, 
which is suggestive of a superfluid ground state.

This work was supported in part by a faculty travel grant from the Office of International
Studies at The Ohio State University (WOP), the Swiss National Science Foundation (WOP),
an ITP Scholar position under NSF grant PHY94-07194 (WOP), EPSRC Grant No. GR/L86852 (MUL)
and by NSF grant DMR-9616574 (RRPS). We thank the ETH-Z\"urich (WOP) and the ITP at
UCSB (WOP, RRPS) for hospitality while this work was being completed.\bigskip

\noindent
$^*$Permanent address.

\end{document}